\documentclass[pra,twocolumn,showpacs,a4paper]{revtex4-1}
\usepackage{epsfig}
\usepackage{amsmath}
\usepackage{graphicx}
\usepackage{bbm}

\begin{document}

\title{The photon blockade effect in optomechanical systems}
\author{P. Rabl}
\affiliation{Institute for Quantum Optics and Quantum Information of the  Austrian Academy of Sciences, 6020 Innsbruck, Austria}

\date{\today}

\begin{abstract}
We analyze the photon statistics of a weakly driven optomechanical system and discuss the effect of  photon blockade under single photon strong coupling conditions. 
We present an intuitive interpretation of this effect in terms of displaced oscillator states and derive analytic expressions for the cavity excitation spectrum and the two photon correlation function $g^{(2)}(0)$. Our results predict the appearance of non-classical photon correlations in the  combined strong coupling and sideband resolved regime, and provide a first detailed understanding of photon-photon interactions in strong coupling  optomechanics.
\end{abstract}

%

         
\maketitle


The implementation of strong optical non-linearities on a single photon level is one of the central goals in quantum optics with a significant practical relevance for applications ranging from optical computation \cite{GibbsBook} to quantum information processing~\cite{BouwmeesterQIPBook} and photonic quantum simulation schemes~\cite{QSimulation}. The prototype system that has been widely studied in this context is cavity QED~\cite{BermanCQEDBook} 
where under strong coupling conditions effective photon non-linearities result from the hybridization between the optical field and a  single atom.
%
Recently, a fundamentally different type of light-matter interaction has attracted a lot of attention, which is the radiation pressure coupling between light and mechanical motion  studied in optomechanical systems (OMS)~\cite{NJPFocusedIssueOM}. 
In most experiments today radiation pressure forces are fairly weak and non-linear optical effects~\cite{MeystreOSAB1985,MarquardtPRL2006} occur in the classical, high photon number regime, where enhanced but linear photon-phonon interactions~\cite{GroblacherNature2009,TeufelNature2011}\ are investigated for cooling~\cite{GroundStateCoolingExp} or the preparation of mechanical quantum states~\cite{AkramNJP2010}.
However, strong optomechanical interactions with \emph{single} photons in analogy to cavity QED, are within reach of new generations of nano-fabricated OMS~\cite{EichenfieldNature2009} or superconducting devices~\cite{TeufelNature2011} and are already nowadays accessible in analogous cold atom experiments~\cite{GuptaPRL2007,BrenneckeScience2008}.  This could open up a new route towards non-linear quantum optics, which avoids single atom strong coupling and where instead photon-photon interactions are mediated by the motion of a macroscopic object.

\begin{figure}
\begin{center}
\includegraphics[width=0.45\textwidth]{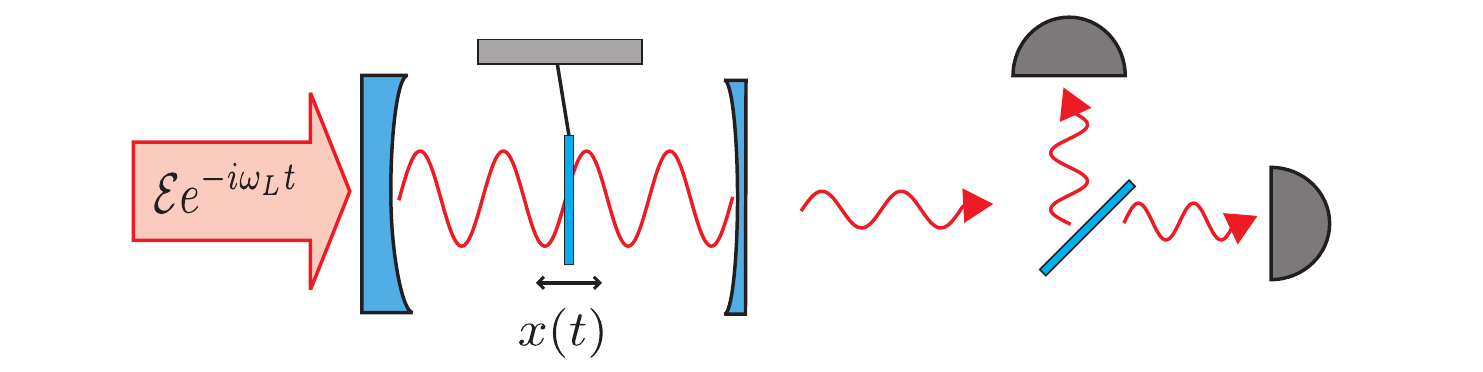}
\caption{Setup for the detection of photon blockade effects in optomechanical systems (OMS). The OMS is weakly excited by a coherent laser field and the statistics of the output field is inferred from photon coincidence measurements.
}
\label{fig:Setup}
\end{center} 
\end{figure}

In this work we study OMS in the regime where the single photon coupling $g_0$ is comparable with the cavity decay rate $\kappa$. In contrast to previous studies~\cite{ManciniPRA1997,BosePRA1997,MarshallPRL2003,LudwigNJP2008,RodriguesPRL2010} we here  focus explicitly on the consequences of strong coupling for the quantum statistics of light, with the aim to identify the mechanism for photon-photon interactions in this system and the experimental conditions under which such effects could be observed in experiments.
To do so we consider a weakly driven OMS as shown in Fig.~\ref{fig:Setup} and evaluate the two photon correlation function $g^{(2)}(0)$. This quantity provides  a direct experimental measure for non-classical anti-bunching effects, i.e.  $g^{(2)}(0)<1$, and the limit $g^{(2)}(0)\rightarrow 0$ indicates a complete \emph{photon blockade}~\cite{TianPRA1992,ImamogluPRL1997,BirnbaumNature2005}, where strong photon-photon interactions prevent multiple photons from passing through the cavity at the same time.
%
%
Our analysis shows that apart from $g_0$  
signatures of non-classical light in OMS depend crucially on the relation between the cavity linewidth $\kappa$ and the mechanical frequency $\omega_m$ and appear only under quite stringent conditions $\kappa<  g_0, \omega_m$. However, this regime is within reach of experiments~\cite{EichenfieldNature2009,TeufelNature2011,GuptaPRL2007,BrenneckeScience2008} where the observation of photon blockade would provide the essential ingredient for potential applications of OMS as a quantum non-linear device.

\emph{Model.} We consider a setup as shown in Fig.~\ref{fig:Setup}, where the frequency of an optical cavity mode is modulated  by the linear displacement of a mechanical oscillator. The cavity is excited by a weak coherent laser field and 
the photon statistics of the transmitted light is analyzed using standard photon counting techniques~\cite{BirnbaumNature2005}.  
In a frame rotating with the laser frequency $\omega _{L}$ the Hamiltonian for the OMS is $(\hbar=1)$
\begin{equation}\label{eq:Hop}
H_{op}= H_m -\Delta _{0}  c^\dag c + \sum_k g_k(b^\dag_k+b_k) c^\dag c  +i  \mathcal{E} \left(c^\dag-c\right),
\end{equation}
where  $c$ is the bosonic operator for the cavity mode,  $\mathcal{E}$ is the driving strength and $\Delta_0=\omega_L-\omega_c$ is the detuning of the laser from the bare cavity frequency $\omega_c$. 
The bosonic operators  $b_k$ represent the mechanical eigenmodes of system which evolve under the free Hamiltonian $H_m=\sum_k \omega_k b^\dag_k b_k$ and couple to the cavity with a strength $g_k$.  In a general device the $b_k$'s  account for different vibrational modes of the resonator as well as mechanical modes of the support  
and the system specific details of the OM interactions are summarized by the spectral density $J(\omega)=\frac{\pi}{2}\sum_k g_k^2 \delta(\omega-\omega_k)$~\cite{LeggettRMP1987}. 
For concreteness we will below focus  explicitly on the case
\begin{equation}\label{eq:J}
J(\omega)=  \frac{\omega}{Q} \frac{\eta^2}{(\omega^2/\omega^2_m-1)^2+ \omega^2/(\omega_m^2Q^2)},
\end{equation}
which models a mechanical mode of frequency $\omega_m$ coupled to an Ohmic bath. The dimensionless parameter  $\eta=g_0/\omega_m$ is chosen such that in the limit of a high mechanical quality factor $Q$  we recover the standard model for a single mode OMS~\cite{NJPFocusedIssueOM} with a  coupling constant $g_0$.

The cavity field is coupled to the electromagnetic vacuum modes of the environment and  in the limit of a single sided cavity we model the resulting dissipative dynamics by a quantum Langevin equation
\begin{equation}\label{eq:QLE} 
\dot c(t)=i[H_{op},c(t)]-\kappa c(t) -\sqrt{2\kappa} f_{in}(t).
\end{equation}
Here $\kappa$ is the cavity field decay rate and $f_{in}(t)$ a $\delta$-correlated noise operator. Photon counting and photon coincidence measurements of  the cavity output field $f_{out}(t)=f_{in}(t)+\sqrt{2\kappa}c(t)$ provide information about the photon statistics of the cavity field~\cite{BirnbaumNature2005}. 


\begin{figure}
\begin{center}
\includegraphics[width=0.45\textwidth]{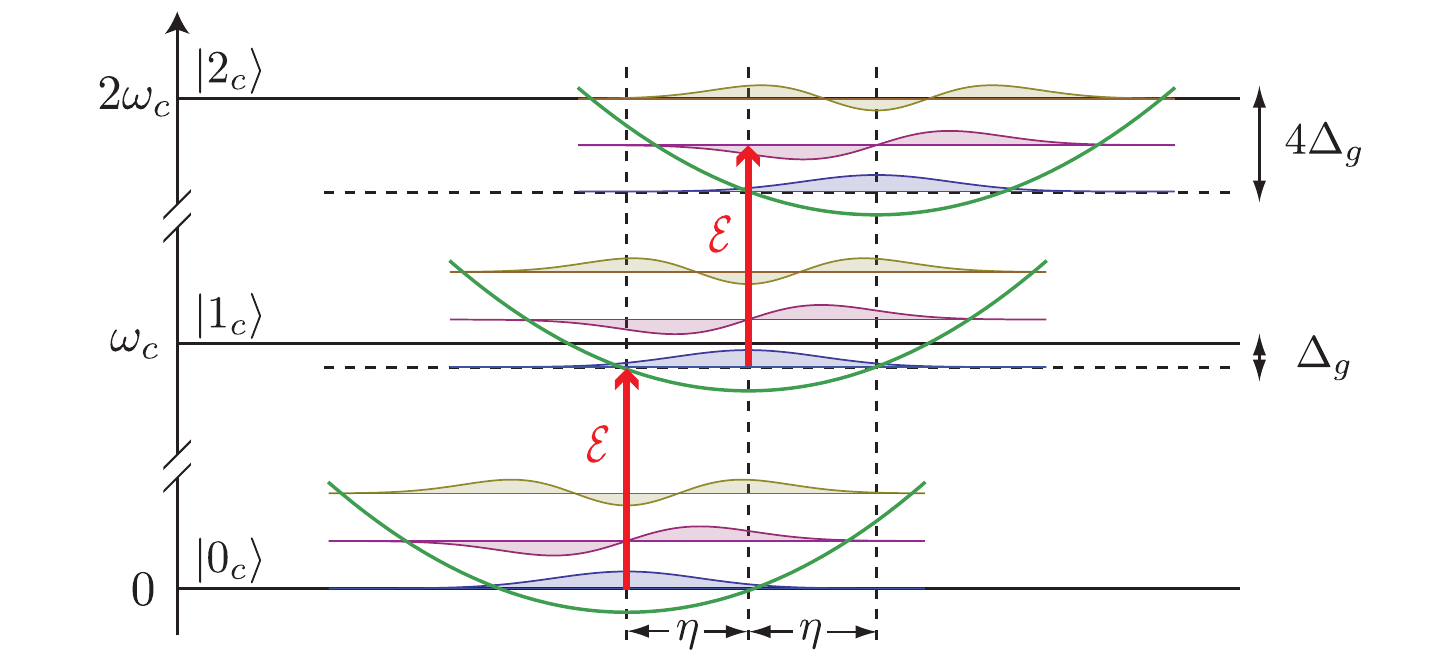}
\caption{Level diagram of the isolated, single mode OMS where $\eta=g_0/\omega_m$ and $\Delta_g=g_0^2/\omega_m$. For a photon number state $|n_c\rangle$ the radiation pressure displaces the resonator equilibrium  $\sim n_c\times \eta$. As a result the energy of the photon states is lowered by $n_c^2\times \Delta_g$ and leads to different resonance conditions for the first  and the second  laser photon exciting the cavity.}
\label{fig:Displaced}
\end{center} 
\end{figure}

\emph{Displaced oscillator states.}   
For the observation of photon blockade effects 
we are interested in the regime of low photon numbers where the driving field $\mathcal{E}$ only weakly perturbs the OMS.
Therefore,  to proceed it is convenient to change to a displaced oscillator representation, $H_{op} \rightarrow UH_{op}U^\dag$, which diagonalizes $H_{op}$ in the limit $\mathcal{E}\rightarrow 0$ and is defined by
the unitary transformation  $U=e^{-iP c^\dag c}$ and $P=i\sum_k (g_k/\omega_k)(b_k^\dag -b_k) $.
We obtain
\begin{equation}\label{eq:Htilde}
 H_{op}= H_m-\Delta  c^{\dag }c - \Delta_g c^\dag c^\dag c c +i  \mathcal{E} \left(c^\dag e^{-iP} - e^{iP} c  \right),
\end{equation}
where we have introduced a shifted detuning $\Delta=\Delta_0+\Delta_g$ and a photon non-linearity 
$
\Delta_g=\frac{2}{\pi} \int_0^\infty d\omega J(\omega)/\omega
$, where  $\Delta_g=g_0^2/\omega_m$ for the single mode model defined in Eq.~\eqref{eq:J}.  
The origin of this effective photon-photon interaction can be understood from the fact that in an isolated system the radiation pressure force displaces the resonator equilibrium  by an amount proportional to the photon number $n_c$ and thereby lowers the energy of this photon state by $n_c^2\times \Delta_g$. 
This is illustrated in more detail in Fig.~\ref{fig:Displaced} and already explains the basic mechanism for photon blockade. If the driving laser is on resonance with the $|0_c\rangle \rightarrow|1_c\rangle$  transition, i.e. $\Delta=0$, the same $|1_c\rangle\rightarrow|2_c\rangle$ transition is detuned by $2\Delta_g$ and will be suppressed for  $\Delta_g> \kappa$. 
However, this simple picture is based on the level structure of the isolated OMS~\cite{ManciniPRA1997,BosePRA1997} only and 
ignores phonon sideband transitions and other dynamical  aspects of the problem which  will be addressed by the following more rigorous analysis.

\emph{Excitation spectrum.} We first study the cavity excitation spectrum  $S(\Delta_0):=\lim_{t\rightarrow\infty}  \langle c^\dag(t) c(t)\rangle/n_0$, which is normalized to the resonant  photon number of the unperturbed cavity,  $n_0=\mathcal{E}^2/\kappa^2$.  The OMS is initially prepared in the state $\rho(0)=|0_c\rangle\langle 0_c|\otimes \rho_{th}$ where $\rho_{th}$ is the thermal equilibrium state of the mechanical modes. 
For a finite, but weak driving field the dominant contribution for $S(\Delta_0)$ arises from terms in the Heisenberg operator $c(t)$ which are linear in $\mathcal{E}$ and from  the displaced oscillator representation of Eq.~\eqref{eq:QLE} we obtain  
\begin{equation}\label{eq:cdot}
\dot c(t)= (i\Delta-\kappa) c(t) +  e^{-iP(t)} (\mathcal{E} -\sqrt{2\kappa}f_{in}(t)) + \mathcal{O}(\mathcal{E}^2).
\end{equation}
On the same level of accuracy the operator $P(t)$ can be approximated by  the free evolution $P(t)=e^{-iH_m t}Pe^{iH_m t}$ 
and  after integrating Eq.~\eqref{eq:cdot} we find
\begin{equation}\label{eq:G1}
S(\Delta_0)= \kappa \, {\rm Re} \int_{0}^{\infty} d\tau \, e^{(i \Delta-\kappa)\tau} e^{-F_2(\tau)} +\mathcal{O}(\mathcal{E}).
\end{equation} 
Here $e^{-F_2(\tau)}=\langle e^{iP(\tau)}e^{-iP(0)} \rangle$ is the equilibrium correlation function of the mechanical displacement operator~\cite{LeggettRMP1987}. 
For later convenience we define  $f_2\equiv f_2(\omega,\tau)=1-e^{-i\omega \tau}$ and write the function $F_2(\tau)$ in the general form 
\begin{equation}\label{eq:Fn}
F_k(\{\tau_i\})=  \frac{2}{\pi}\int_0^\infty  d\omega \frac{J(\omega)}{\omega^2} \left[ (N(\omega)\!+\!1)f_k\!+\! N(\omega) f_k^*\right],
\end{equation}
where $N(\omega)=1/(e^{\hbar \omega/k_BT}-1)$ is the bosonic equilibrium occupation number for a support temperature $T$.

\begin{figure}
\begin{center}
\includegraphics[width=0.45\textwidth]{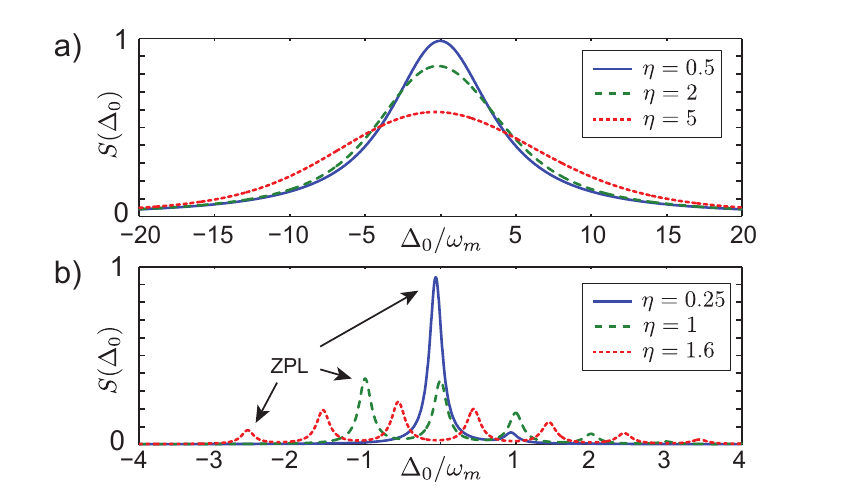}
\caption{Cavity excitation spectrum $S(\Delta_0)$ for different values of the coupling parameter $\eta=g_0/\omega_m$ and a) $\kappa/\omega_m=4$ and b) $\kappa/\omega_m=0.1$. In b) the zero phonon line (ZPL) indicates the position of the phonon number conserving transition. In both plots $T=0$ and $Q=150$.  }
\label{fig:ExcitationSpectrum}
\end{center} 
\end{figure}

 Fig.~\ref{fig:ExcitationSpectrum} shows the cavity excitation spectrum $S(\Delta_0)$ for two different values of $\kappa/\omega_m$ and for $J(\omega)$ defined in Eq.~\eqref{eq:J}.   In the bad cavity limit $\kappa\gg \omega_m$ we can approximate $F_2(\tau)\simeq i\Delta_g \tau +  \tau^2/(4T_\varphi^2)$  and we identify an optomechanical dephasing mechanism with a  timescale $T_\varphi^{-1}=2g_0\sqrt{N +1/2} $ where $N:=N(\omega_m)$.
This leads to a broadening of the spectrum and a gradual change from a Lorentzian to a Gaussian lineshape  $ S(\Delta_0)\approx  \sqrt{\pi} \kappa T_\varphi  e^{-\Delta_0^2T_\varphi^2}$ for very large values of $g_0$. A completely different behavior is found in the sideband resolved regime $\kappa \ll \omega_m$. Here we observe a red-shift of the zero phonon line (ZPL) towards $\Delta_0=-\Delta_g$ and the appearance of additional  resonances at multiples of the mechanical frequency $\omega_m$.  These peaks result from phonon assisted excitation processes. 

For a more detailed quantitative discussion of $S(\Delta_0)$ we now focus on the limit $Q\gg 1$ of a weakly damped mechanical mode. In this regime  $F_2(\tau)=F_2^r(\tau)+iF_2^i(\tau)$, where $ F_2^i\simeq \eta^2 \sin(\omega_m \tau) e^{-\frac{\gamma}{2} \tau} $ and 
 $
 F_2(\tau)\simeq \Gamma \tau + \eta^2 (2N+1) (1\!-\cos(\omega_m \tau) e^{-\frac{\gamma}{2} \tau}).
$
Here $\gamma=\omega_m/Q$ is the mechanical damping rate and $\Gamma$ is an additional decoherence rate which arises from the low frequency part of $J(\omega)$. It vanishes for $T\rightarrow 0$ and is  given by $\Gamma \simeq \eta^2 (2N\!+\!1)\gamma$ for temperatures $T\geq \hbar \omega_m/k_B$.  The approximate analytic expression for $F_2(\tau)$ allows us to expand the two point correlation function in Eq.~\eqref{eq:G1} and evaluate the integral over $\tau$~\cite{HuangRhys}. We obtain
\begin{equation}\label{eq:ExSpec}
S(\Delta_0)\simeq   \kappa\,  \sum_{n=-\infty}^\infty  A_n \frac{\kappa_n}{\kappa_n^2+(\Delta_0+\Delta_g-n \omega_m)^2},
 \end{equation}
where  $A_n\!=\! e^{-\eta^2 (2N\!+\!1) }    I_n(2\eta^2\sqrt{ N(N\!+\!1)} ) \left( (N\!+\!1)/N \right)^\frac{n}{2}$ and $I_n(x)$ is the $n$-th order modified Bessel function.  This result is familiar from the standard Huang-Rhys theory of phonon assisted excitation processes~\cite{HuangRhys} and the positions and  weights of the resonances can be understood from different multi-phonon sidebands of the $|0_c\rangle\rightarrow|1_c\rangle$ transition shown in Fig.~\ref{fig:Displaced}.
Apart from photon loss the resonances are broadened by the mechanical decoherence rates where $\kappa_n\simeq\kappa +\Gamma+|n|\gamma/2$ for $N\lesssim  1$ and $\kappa_n\simeq\kappa+2\Gamma$ in the hight temperature limit.
We emphasize  that at $T= 0$ both the appearance of  phonon sidebands for $\kappa< \omega_m$ as well as the broadening of the cavity resonance in the opposite regime  $\kappa> \omega_m$
are pure quantum effects and provide a clear indication for single photon strong coupling optomechanics.  

\emph{Photon correlations.} For weak driving the excitation spectrum is dominated by single photon events and does not contain information about photon-photon interactions. To proceed we now concentrate on the normalized equal time correlation function $g^{(2)}(0):= \lim_{t\rightarrow \infty} \langle c^\dag c^\dag cc\rangle(t)/ \langle c^\dag c\rangle^2(t)$, where in addition to $S(\Delta_0)$ we must evaluate  
the two photon correlation $G^{(2)}:=\lim_{t\rightarrow\infty}  \langle c^{\dag 2}(t) c^2(t)\rangle/n^2_0$. Following the same arguments as above we obtain up  to the  lowest relevant order in $\mathcal{E}$,
\begin{equation}
\dot c^2(t)= 2(i\Delta +i\Delta_g - \kappa)c^2(t) +\mathcal{E} e^{-iP(t)} c(t)+ \mathcal{O}(\mathcal{E}^3),
\end{equation}
where we have already omitted an irrelevant noise term $\sim f_{in}(t)$.  Together with Eq.~\eqref{eq:cdot} we finally obtain
 \begin{equation}\label{eq:G2}
\begin{split}
&G^{(2)} = 2\kappa^3 \,{\rm Re}\int_{0}^{\infty} d\tau_1  \int_0^{\infty} d\tau_2  \int_0^{\infty} d\tau_3 \\
  \times &e^{2(i \Delta+i\Delta_p-\kappa)\tau_1}  e^{(-i\Delta-\kappa) \tau_2}  e^{(i\Delta-\kappa) \tau_3} e^{-F_4(\tau_1,\tau_2,\tau_3)}, 
 \end{split}
\end{equation}
where $e^{-F_4(\{\tau_i\})}= \langle e^{iP(\tau_1-\tau_2)} e^{iP(\tau_1)} e^{-iP(0)} e^{-iP(-\tau_3)}\rangle$ is a four point correlation function of the mechanical displacement operator. The function $F_4(\{\tau_i\})$ can be expressed in terms of Eq.~\eqref{eq:Fn} by setting $f_4\equiv f_4(\omega,\{\tau_i\})= 2+e^{i\omega \tau_2}+e^{-i\omega \tau_3} - 
  (1+e^{i\omega\tau_2}) e^{-i\omega \tau_1}(1+e^{-i\omega \tau_3})$.

Since a general discussion of  Eq.~\eqref{eq:G2} is quite involved we will from now on  concentrate on the most relevant regime where the mechanical decoherence rates $\Gamma$ and $\gamma$ can be neglected compared to $\kappa$. 
However, we first point out that in the bad cavity limit we can approximate  $F_4(\{\tau_i\})\approx i\Delta_p (4\tau_1 - \tau_2 + \tau_3)+ \left(2\tau_1-\tau_2+\tau_3\right)^2/(4T_\varphi^2)$. Then, for  $\kappa,\Delta_0 <T_\varphi^{-1}$ we obtain $g^{(2)}(0)\approx  e^{(\Delta_0T_\varphi)^2} /(\sqrt{4\pi} \kappa T_\varphi) >1$ and  we conclude 
that even for strong coupling $g_0$ the photon statistics of an OMS in the bad cavity limit remains classical.    

%
Let us now consider the limit $Q\rightarrow \infty$ where 
$F_4(\{\tau_i\}) \simeq \eta^2 f_4(\omega_m,\{\tau_i\})$ and as above, we use a series expansion of the correlation function in Eq.~\eqref{eq:G2} to evaluate  the integrals over the  $\tau_i$. We obtain 
 \begin{equation}\label{eq:G2Sum}
\begin{split}
G^{(2)} = {\rm Re} \!\sum_{n,m,p}  &\!  \frac{B_{n,m,p}}{(\kappa+i(\Delta\!-\!n \omega_m))(\kappa-i(\Delta\!-\!m \omega_m))}\\
\times  &  \frac{2\kappa^3}{(2\kappa-i(2\Delta\!+\!2\Delta_g\!-\!p \omega_m))},
\end{split}
\end{equation}
where the coefficients $B_{n,m,p}$ depend on the coupling parameter $\eta$ and follow from,
\begin{equation}
e^{-\eta^2 f_4(\omega_m,\{\tau_i\})}= \sum_{n,m,p} B_{n,m,p} \,e^{i\omega_m (\tau_2 n-\tau_3 m-\tau_1 p)}.
\end{equation}
For example,  for $T=0$ explicit expressions are given by 
$B_{n,m,p} = e^{-2\eta^2} (\eta^2)^p  W_{n,p}(\eta)W_{m,p}(\eta)/n!m!p!$
where $W_{n,p}(\eta)=(-1)^n{\rm U}[-n,1-n+p,\eta^2]$ and ${\rm U}[a,b,x]$ is a confluent hypergeometric function.

  \begin{figure}
\begin{center}
\includegraphics[width=0.45\textwidth]{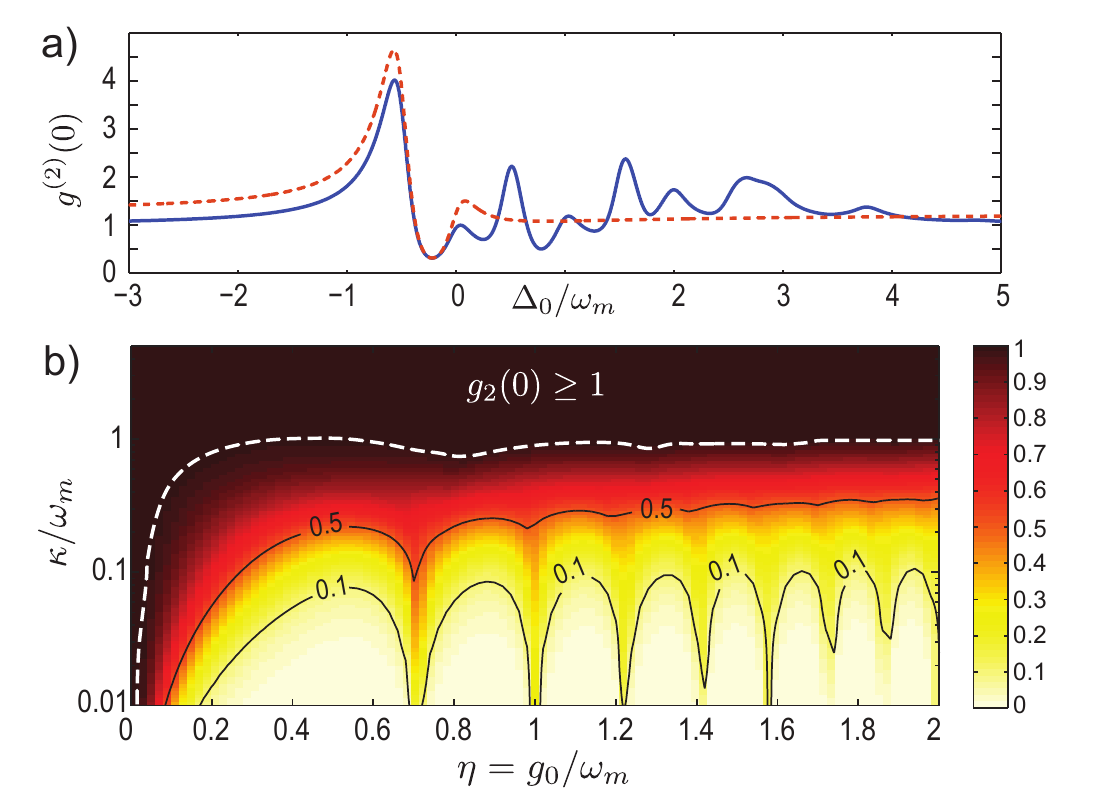}
\caption{a) Dependence of $g^{(2)}(0)$ on the bare laser detuning $\Delta_0$ for $g_0/\omega_m= 0.5 $ and $\kappa/\omega_m=0.15$. The dashed line indicates the approximate result given in Eq.~\eqref{eq:Appg2}.  b) The minimum of $g^{(2)}(0)$ with respect to $\Delta_0$ is plotted for different values of $\kappa$ and $g_0$. In both plots  $T=0$ and $Q\rightarrow \infty$.  }
\label{fig:G2Plot}
\end{center} 
\end{figure}

\emph{Discussion.} In Fig.~\ref{fig:G2Plot} a) we use Eq.~\eqref{eq:ExSpec} and Eq.~\eqref{eq:G2Sum} to evaluate $g^{(2)}(0)$ and plot the result as a function of  $\Delta_0$ and for $\kappa \ll \omega_r$.  We observe a  sequence of bunching and anti-bunching resonances which can be qualitatively understood from the level diagram shown in Fig.~\ref{fig:Displaced}, where depending on  $\Delta_0$
 either the one or the two photon transition becomes resonant with different phonon sidebands. For a better understanding of this process we now consider the regime $\eta<1$ and assume that the laser is tuned close to the ZPL of the one photon transition, i.e. $\Delta=\Delta_0+\Delta_g\ll \omega_m$. 
Then, still assuming $\kappa < \omega_m$, the dominant contributions to the sum in Eq.~\eqref{eq:G2Sum} arise from the terms $n=m=0$ and $p=0,1$ and combined with the $n=0$ terms in Eq.~\eqref{eq:ExSpec} we obtain 
  \begin{equation}\label{eq:Appg2}
g^{(2)}(0) \simeq   \left[  \frac{C_0(\kappa^2+\Delta^2)}{\kappa^2\!+\!(\Delta\!+\!\Delta_g)^2}\!+\! \frac{\eta^2 C_1 (\kappa^2+\Delta^2)}{\kappa^2\!+\!(\Delta\!+\!\Delta_g\!-\! \omega_m/2)^2}\right].
\end{equation}
Here $C_0=B_{0,0,0}/A_0^2$ and $C_1=B_{0,0,1}/(\eta^2A_0^2)$ and for zero temperature $C_{0}=C_{1}=1$. 
Fig. \ref{fig:G2Plot} a) shows that Eq.~\eqref{eq:Appg2} provides indeed an excellent  approximation of the first anti-bunching tip around $\Delta_0\simeq -\Delta_g$ and allows us to make the following analytic predictions. First, for $\kappa > g_0$ the minimum for $g^{(2)}(0)$ occurs at $\Delta\approx \kappa $ and scales as ${\rm min}\{g^{(2)}(0)\}\approx 1- g_0^2/(\omega_m\kappa)$. Therefore,  as expected, no significant anti-bunching effects appear unless the strong coupling condition $g_0> \kappa$ is achieved. In this regime the minimum of $g^{(2)}(0)$ occurs at $\Delta= 0$ and 
\begin{equation}
{\rm min}\{g^{(2)}(0)\} \simeq    \frac{\kappa^2}{\omega_m^2}\left[ \frac{1}{\eta^4} + \frac{4\eta^2}{(\kappa/\omega_m)^2+(1-2\eta^2)^2 }\right].
\end{equation}
This result demonstrates that OMS can indeed exhibit a strong photon blockade effect where for $\eta\ll 1$ the suppression of two photon events scales with the parameter $\kappa^2 \omega_m^2/g_0^4$. However, rather than improving monotonically with increasing coupling strength the blockade reaches a minimum at $\eta\simeq 0.5$ with a value  $g^{(2)}(0)\approx 20 \times (\kappa/\omega_m)^2$.  This minimum is a consequence of the $|1_c\rangle\rightarrow|2_c\rangle$ transition getting into resonance with the first phonon sideband which occurs for $\eta=1/\sqrt{2}$. Therefore, the fidelity of the photon blockade effect is ultimately limited by the sideband parameter $\kappa/\omega_m$ and the effect vanishes as this parameter approaches one.  


Finally, we use Eqs.~\eqref{eq:ExSpec}  and~\eqref{eq:G2Sum} to evaluate the minimum of $g^{(2)}(0)$ for a large range of parameters $g_0$ and $\kappa$ numerically. The results are plotted in Fig.~\ref{fig:G2Plot} b) and show a clear boundary at $\kappa\simeq \omega_m$ which -- quite independently of the value of $g_0$ -- separates the regimes of classical and quantum correlations. We also see that in the sideband resolved regime $\kappa< \omega_m$  the photon blockade exhibits a repetitive pattern and  for $g_0>\omega_m/2$ no significant further improvement is achieved. These results show that the approximate result~\eqref{eq:Appg2} already captures the essence of the two photon blockade effect in OMS.

In summary we have identified the mechanism for strong photon-photon interactions in OMS and studied the dependence of the two photon blockade effect on the relevant parameters $g_0$, $\kappa$ and $\omega_m$.  Our results provide a guideline for future experiments and a first detailed theoretical description of the two photon physics which is relevant for applications of OMS in the context of quantum information processing or  quantum simulation.


\paragraph*{Acknowledgments.} The author thanks D. Stamper-Kurn, K. Hammerer, I. Wilson-Rae, W. Zwerger and P. Zoller  
for stimulating discussions and acknowledges support by the NSF through a grant for ITAMP.

\emph{Note.} After submission of this manuscript a related work by A. Nunnenkamp {\it et al.} appeared on the arXiv:1103.2788.

\end{document}